\documentclass[pre,aps,twocolumn,showpacs]{revtex4}
\usepackage{graphicx}

\begin{document}

\title{Electronic transport in single-helical protein molecules: Effects of multiple charge 
conduction pathways and helical symmetry}

\author{Sourav Kundu}

\email{sourav.kunduphy@gmail.com}

\affiliation{Condensed Matter Physics Division, 
Saha Institute of Nuclear Physics, 
1/AF, Bidhannagar, Kolkata 700 064, India}

\author{S. N. Karmakar}

\affiliation{Condensed Matter Physics Division, 
Saha Institute of Nuclear Physics, 
1/AF, Bidhannagar, Kolkata 700 064, India}

\begin{abstract}

We propose a tight-binding model to investigate electronic transport 
properties of single helical protein molecules incorporating both the 
helical symmetry and the possibility of multiple charge transfer pathways. 
Our study reveals that due to existence of both the multiple charge 
transfer pathways and helical symmetry, the transport properties are 
quite rigid under influence of environmental fluctuations which indicates 
that these biomolecules can serve as better alternatives in nanoelectronic 
devices than its other biological counterparts {\it e.g.}, single-stranded DNA.

\end{abstract}

\pacs{72.15.Rn, 73.23.-b, 73.63.-b, 87.14.gk} 

\maketitle

\section{Introduction}

Current days biomolecules are receiving huge attention from different 
scientific communities including physics, chemistry and others because of 
their possible applications in nanoelctronics and the need of understanding 
electronic (spin) transfer process in biological systems~\cite{endres,zutic,genereux,cordes}. 
DNA is one of them which has attracted major attention from the beginning 
of the last decade~\cite{kelley,fink,porath,cai,tran}. Whereas other biomolecules 
such as proteins are less attended in this respect. However with the recent 
progress in chiral-induced spin selectivity (CISS) both DNA and protein are 
getting similar attraction~\cite{gohler,xie,guo_prl,guo_prb1,guo_pnas,galperin, 
eremko,yeganeh,guti,senthil,gersten,mishra,ben} across various disciplines 
as they both have helical structures which can be used for efficient spin polarization. 
In 2011 G{\"o}hler {\it et al}.~\cite{gohler} showed that double-stranded DNA (ds-DNA) 
can be used as a good spin filtering agent with length dependent spin polarization up 
to 60$\%$. Whereas no spin polarization was achieved for single-stranded DNA (ss-DNA). 
These findings are then theoretically supported by Guo {\it et al}.~\cite{guo_prl}. But 
recent experiments suggest that $\alpha$-helical proteins are also quite efficient in spin 
polarization process though it has single helical structure~\cite{mishra,ben}. These results 
open up an opportunity to examine these single-helical structures from a new aspect, different 
models are also proposed to explain these experimental results~\cite{guo_pnas}. In respect of 
electronic transport properties DNA is widely studied, though there are still controversies 
over different experimental results~\cite{porath,cai,tran,zhang,storm,yoo,kasumov}. 
Questions are still remain on reproducibility of the experimental data and underlying charge transfer 
mechanism~\cite{conwell,ratner,dekker,beratan,sourav1,sourav2,sourav3}. Whereas the same 
properties of different protein molecules are less examined. There are few number of reports 
available in literature on the electronic transfer process in proteins~\cite{jin,prytkova,beratan2,
gao,sepunaru} but no such report on the effects of environment on its electronic transport 
properties. It is confirmed by CISS study and related theoretical work that there are multiple 
charge conduction pathways (MCCP) present in single-helical proteins due to which they 
are able to polarize the electron spin~\cite{guo_pnas,mishra,ben}. This possibility of MCCP 
make helical protein molecules very good agents for long-range charge transport. As proteins 
have these MCCP, electrons will face less disturbances/environmental effects during conduction 
and transport characteristics will be much rigid; reproducing experimental results will be much 
simpler with them.

 In this paper we make an attempt to study the electronic transport properties of 
single helical proteins incorporating both the helical symmetry and possibility of MCCP 
within tight-binding framework. We propose a model Hamiltonian to explore electronic 
transport through single-helical proteins and compare our results with another model 
proposed in Ref.~\cite{guo_pnas}. We study different transport properties from localization 
behaviour to I-V response including the effects of environment that are modeled in terms 
of disordered on-site potential of the amino acids within the tight-binding Hamiltonian.
Our investigations show that due to presence of MCCP the effects of environment are much 
smaller which enable long range coherent charge transfer in these biomolecules. Interplay 
of helical symmetry and disorder also has non-trivial effects on localization and I-V 
responses of the protein molecules.

\section{Theoretical Formulation}

\begin{figure}[ht]
\centering

    \includegraphics[width=17mm,height=75mm]{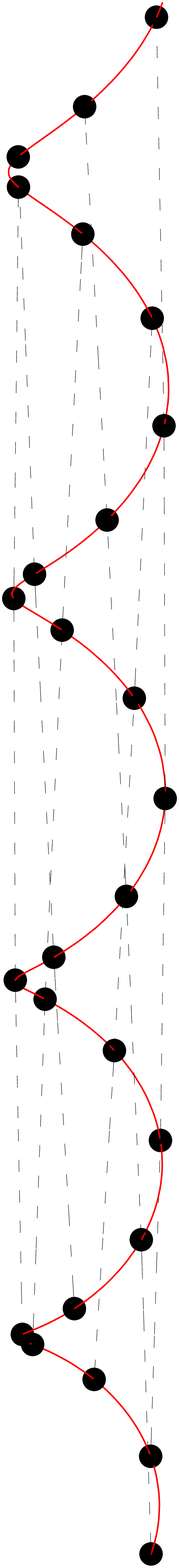}

\caption{(Color online). Schematic diagram of a single-helical protein 
molecule. Black dots on the helix represent the amino acids and dotted 
(black) lines between those black dots represent hopping ($t^{'}$) between 
neighbouring amino acids of adjacent pitches. The red line shows the single 
helix.}

\label{fig1}
\end{figure}

The two-terminal electronic transport through single-helical protein 
molecule can be simulated using the following tight-binding Hamiltonian 
for the entire system 

\begin{equation}
 H_{tot}=H_{pro}+ H_{leads}+H_{tun}~,
\label{hamilton}
\end{equation}
where  $H_{pro}$ is the Hamiltonian for the protein molecule, $H_{leads}$ represents 
the one dimensional semi-infinite leads on the both sides of the protein molecule and 
$H_{tun}$ is the tunneling Hamiltonian between protein molecule and the leads. The 
Hamiltonian~(Fig.\ref{fig1}) for the protein molecule is formed on the basis set spanned 
by the amino acids 
\begin{eqnarray}
& H_{pro}&= \sum\limits_{i=1}^N \epsilon
c^\dagger_{i}c_{i}+ \sum\limits_{i=1}^{N-1} tc^\dagger_{i}c_{i+1}+
\sum\limits_{i=1}^{N-n} t^{'}c^\dagger_{i}c_{i+n}+\mbox{H.c.}~,
\end{eqnarray} 

where $c_i^\dagger$ ($c_i$) is the creation (annihilation) 
operator for electrons at the {\it i}th Wannier state of the 
protein molecule with length N, $t=$ nearest neighbour hopping 
amplitude, $\epsilon=$ on-site potential energy of the amino acids, 
$t^{'}$= hopping integral between two neighboring atomic sites in 
adjacent pitches which incorporates the possibility of MCCP along the 
helix. Here $n$ is the number of amino acids within a given pitch, 
parameter that accounts for the helical symmetry. A dispersion relation 
for an infinite homogeneous chain of protein molecule can be obtained 
following the above Hamiltonian: $E=\epsilon+2t\cos (k)+2t^{'}\cos (nk)$
which explicitly depends on helical symmetry ($n$).

 Apart from the above model, we also use another model following Ref.~\cite{guo_pnas} 
to compare our results. The protein molecule is described in this model 
following the tight-binding Hamiltonian (later on we refer this as model:2 
throughout the paper)

\begin{eqnarray}
& H_{pro}&= \sum\limits_{i=1}^N \epsilon c^\dagger_{i}c_{i}+ 
\sum\limits_{i=1}^{N-1}\sum\limits_{j=1}^{N-i} t_j c^\dagger_{i}c_{i+j}+\mbox{H.c.}~,
\end{eqnarray} 

where $c_i^\dagger$, $c_i$, $\epsilon$ and N have their usual meanings. 
$t_j$= $t_1 e^{(l_j-l_1)/l_c}$ is the jth neighbouring hopping amplitude, 
where $l_j$ is the distance between to neighbour $i$ and $i+j$, $l_c$ is 
the decay exponent and $t_1$ is the nearest neighbour hopping integral. 
Here we have assumed that wave functions related to electrons decay 
exponentially over distance. These assumptions are similar to the 
Slater-Koster scheme, and $l_c$ can be obtained by matching to 
first-principle calculations~\cite{endres,guo_pnas}.

 In order to study the transport behavior of protein molecules, 
we use semi-infinite 1D chains as leads connected to the left 
(L) and right (R) ends of the protein molecule and the corresponding 
Hamiltonian can be expressed as 
 
\begin{eqnarray}
H_{leads}=\sum\limits_{i}\left(\epsilon c^\dagger_ic_i+
t c^\dagger_{i}c_{i+1}+\mbox{H.c.}\right)~,
\end{eqnarray}

where $i<0$ and $i>N$ respectively represent left and right 
semi-infinite 1D leads . The tunneling Hamiltonian between 
the leads and protein molecule is given by $ H_{tun}= \tau \left
(c^\dagger_0c_1+c^\dagger_Nc_{N+1} +\mbox{H.c.}\right)~$
where $\tau$ is the tunneling matrix element between protein and 
the leads.         
	In order to obtain transmission probability $T(E)$ of electron 
through single-helical protein we use the Green's function formalism~\cite{datta1}. 
The single particle retarded Green's function for the entire system at an 
energy $E$ is given by $G^r=(E-H+i\eta)^{-1}$, where $\eta\rightarrow0^+$. 
The transmission probability of an electron with incident energy $E$ is given 
by $T(E)={\mbox {\rm Tr}} [\Gamma_L G^r \Gamma_R G^a]$ where $\mbox {Tr}$ 
represents trace over reduced Hilbert space spanned by the protein molecule.
The retarded and the advanced Green's functions in the reduced Hilbert space 
can be expressed as $G^r=[G^a]^\dagger=[E- H_{pro}-\Sigma^r_L-\Sigma^r_R+i\eta]^{-1}$, 
where $\Sigma^{r(a)}_{L(R)}=H^\dagger_{tun} G^{r(a)}_{L(R)} H_{tun}$ is retarded 
(advanced) self energy of the left (right) lead and 
$\Gamma_{L(R)}=i[\Sigma^r_{L(R)}-\Sigma^a_{L(R)}]$ is the level broadening due 
to coupling of the leads with the protein molecule. $G^{r(a)}_{L(R)}$ being the 
retarded (advanced) Green's function for the left (right) lead. It can easily be 
shown that $\Gamma_{L(R)}=-2~{\mbox{\rm Im}} (\Sigma^r_{L(R)})$, where $\mbox {Im}$ 
represents the Imaginary part. At absolute zero temperature, using the Landauer 
formula, current through the protein molecule for an applied bias voltage $V$ is 
given by $I(V)=\frac{2e}{h} \int^{E_F+eV/2}_{E_F-eV/2} T(E)dE$~, where $E_F$ is 
the Fermi energy. We have assumed that voltage drop occurs only at the 
boundaries of the conductor.

\section{Results:}

To perform numerical study we use following parameter values for our 
proposed model throughout the entire work: $\epsilon$=0 eV, $t$=1.0 eV 
and $t^{'}$=$t$/10=0.1 eV. We compare our results with model:2 using the 
following parameters: $l_1$=4.1, $l_2$=5.8, $l_3$=5.1, $l_4$=6.2, $l_5$=8.9, 
$l_6$=10.0 and $l_c$ is taken as 0.9, all units are in \AA{}. Using these 
values we can calculate the related hopping integrals ($t_j$) which gives 
$t_2$ $\sim$ 0.16$t_1$ and so on. It is clear that gradually $t_j$ values 
will decrease (except $t_3>t_2$) with increasing distance, therefore we 
restrict ourselves to $t_6$ and set $t_1$=$t$=1.0 eV. These parameter values 
for model:2 are extracted from Ref.~\cite{guo_pnas}. For ss-DNA, to calculate 
its transport properties, we set $t^{'}$=0 eV in our model which cancels any 
possibility of MCCP. We first study the localization properties of the system. 
The localization length ($l$) of the system is calculated from the Lyapunov 
exponent ($\gamma$)~\cite{ventra}

\begin{equation}
\gamma = 1/l = -\lim\limits_{N\to\infty}\frac{1}{2N}<\ln(T(E))>~,
\end{equation}

where $N$ = length of the system and $<>$ denotes average over different 
disorder configurations. In actual experiments there are various 
environmental fluctuations that we have simulated in the model by considering 
the on-site energy $\epsilon$ to be randomly distributed within the range 
[$\epsilon$-w/2, $\epsilon$+w/2], where w represents the disorder strength. 

\begin{figure}[ht]
\centering

  \begin{tabular}{c}

    \includegraphics[width=60mm]{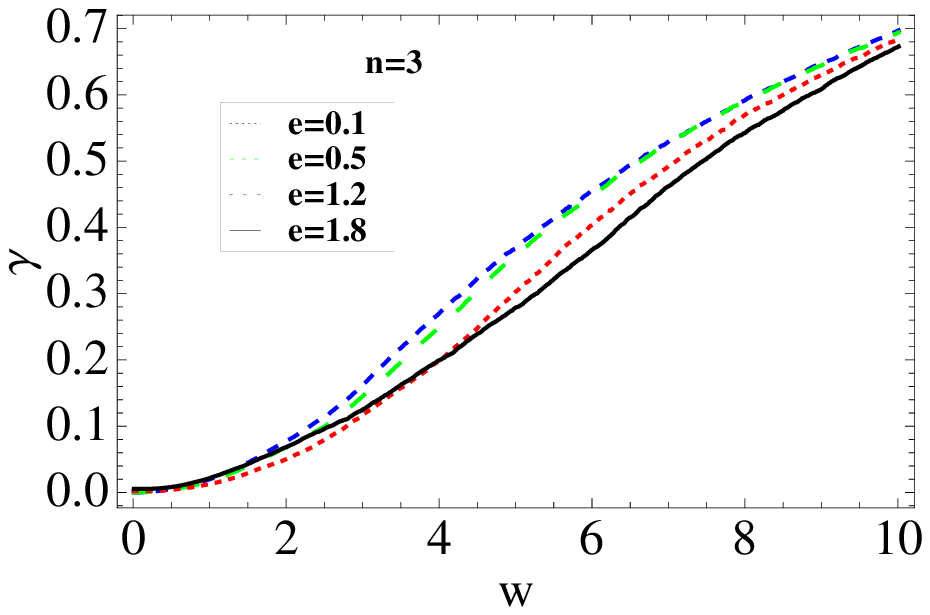}\\
   
    \includegraphics[width=60mm]{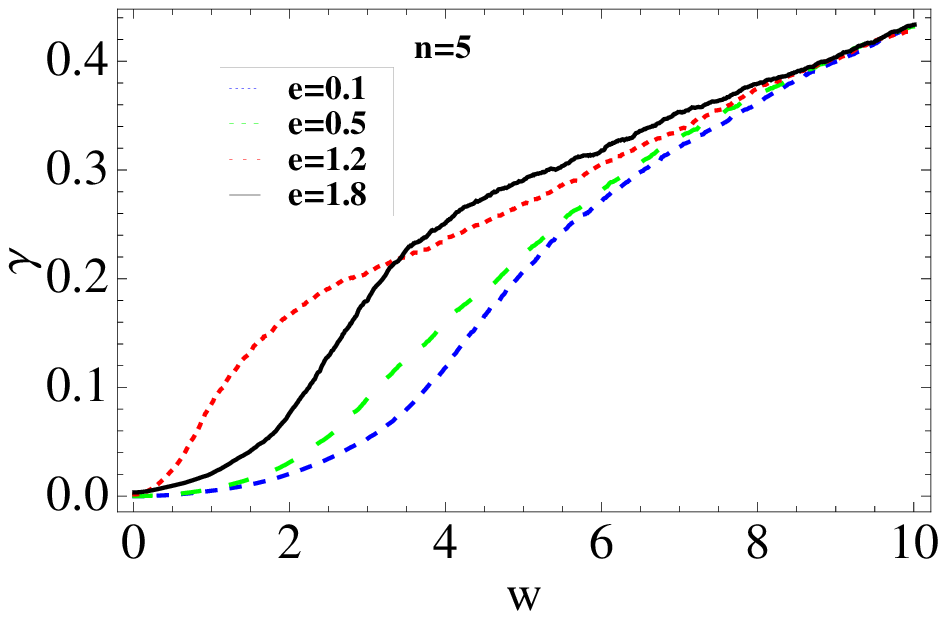}\\

    \includegraphics[width=67mm]{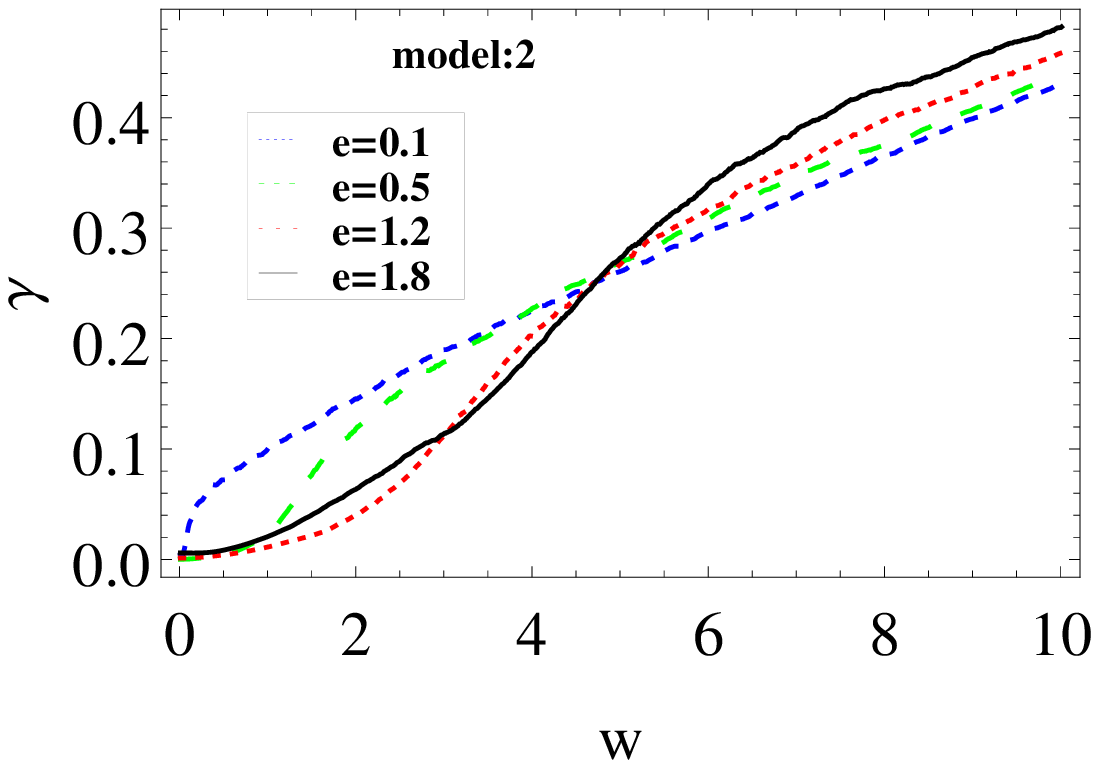}\\
    
    \includegraphics[width=67mm]{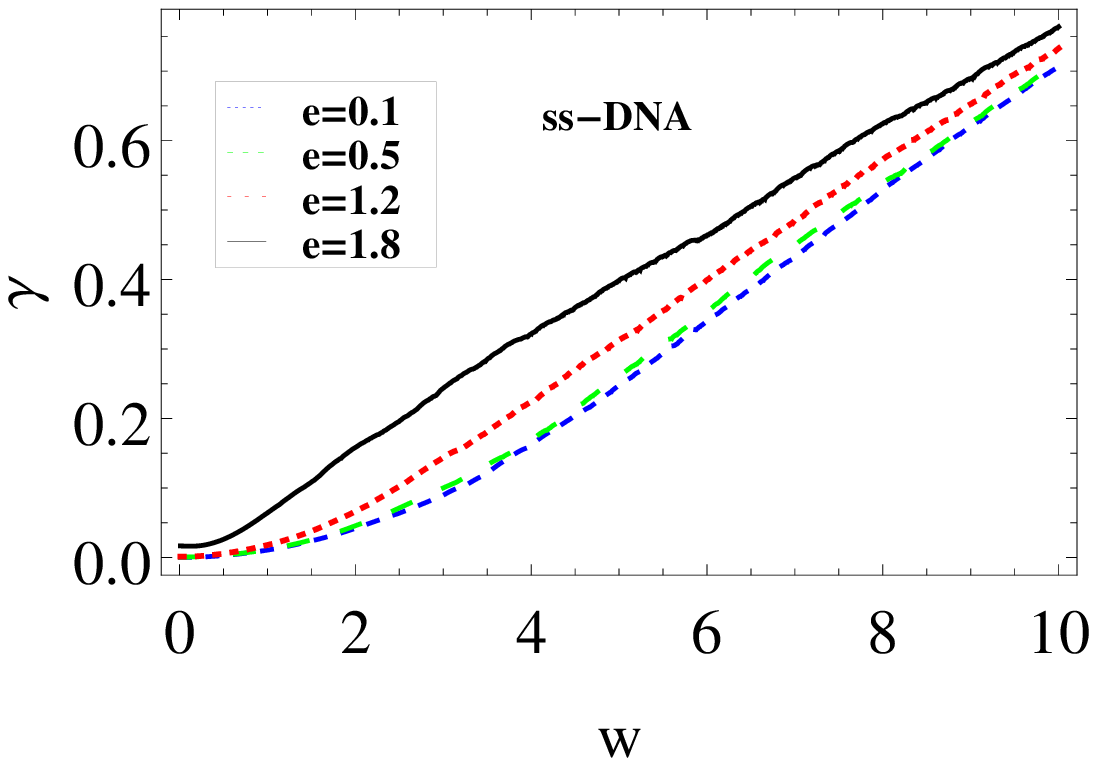}

  \end{tabular}

\caption{(Color online). Lyapunov exponent ($\gamma$) vs disorder (w) 
for $n$=3 and $n$=5, model:2 and for ss-DNA at different energies (e). 
Figure shows that $n$=5 is maximally extended with lowest $\gamma$ 
value for all the energies.}

\label{fig2}
\end{figure}

 In fig.~\ref{fig2} we show the variation of Lyapunov exponent with disorder 
at different energies for both the protein molecule and ss-DNA. For $n$=3, at 
low disorder there is almost no variation in the localization length for different 
energies. With increasing disorder the separation between different energies become 
distinguishable, energies close to the band centre become more localized than the 
band edges. For $n$=5, variation between different energies are maximum at low 
disorder which gradually vanishes with increasing disorder. For model:2, different 
energies show different localization length for almost entire range of disorder 
with a crossover point around w=5. At these disorder strength, localization length 
corresponding to different energies become nearly equal. For ss-DNA, different energies 
remain separated throughout the entire range and the separation between them decreases 
slightly at higher disorder. It is clear form these plots that $n$=5 has the 
maximum localization length for entire disorder range, which shows it is the most 
favourable conformation of the protein molecule for charge transport. 

\begin{figure}[ht]
\centering

  \begin{tabular}{c}

    \includegraphics[width=60mm]{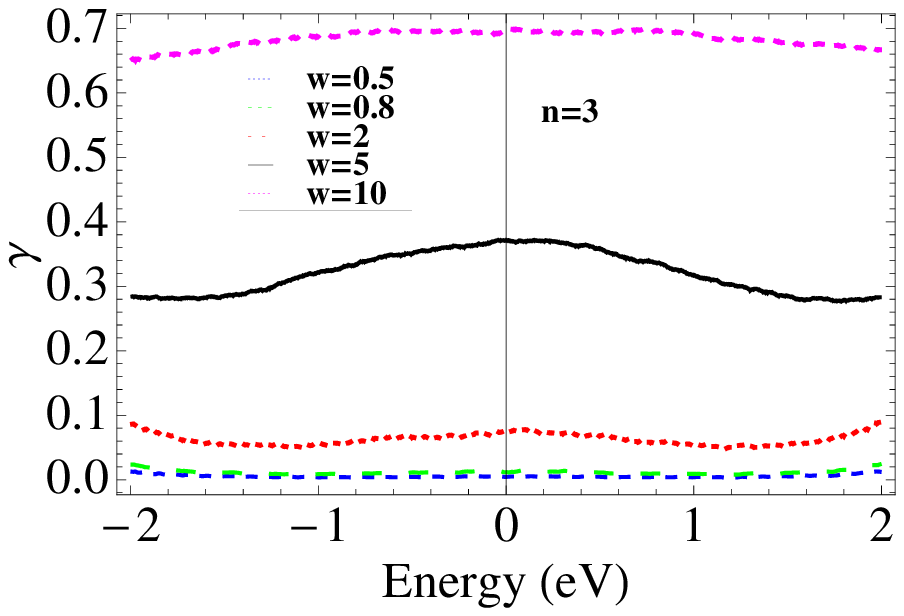}\\
   
    \includegraphics[width=60mm]{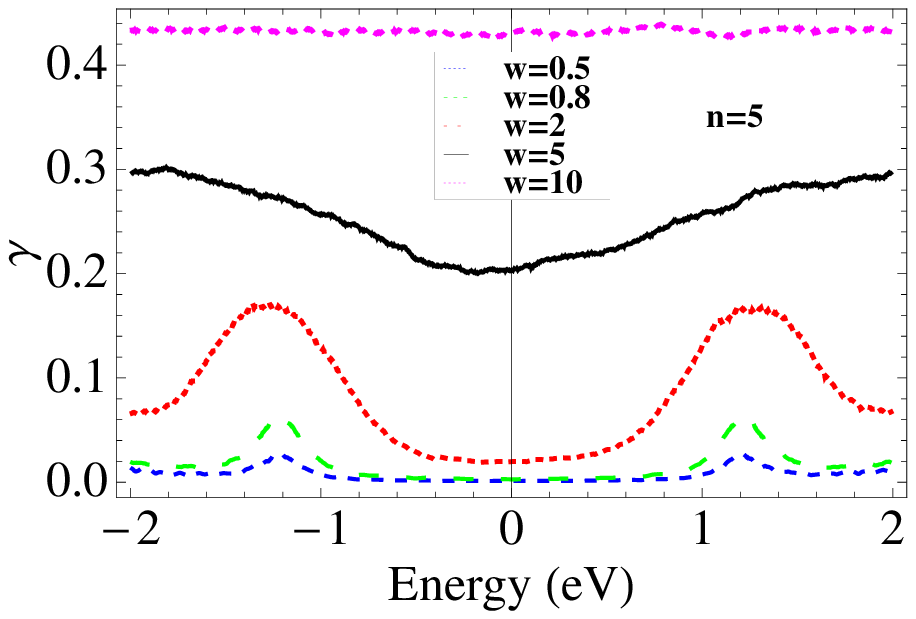}\\
    
    \includegraphics[width=60mm]{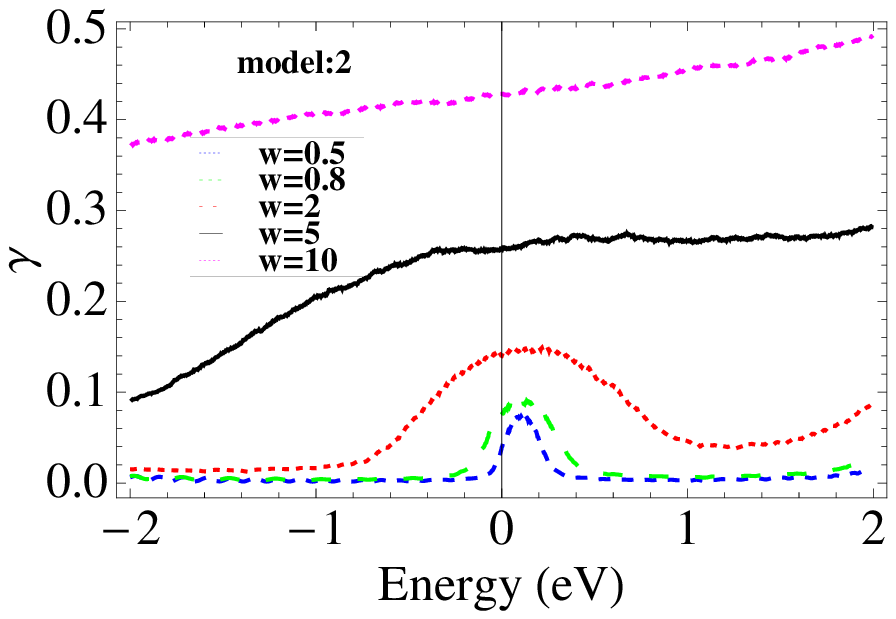}\\
    
    \includegraphics[width=67mm]{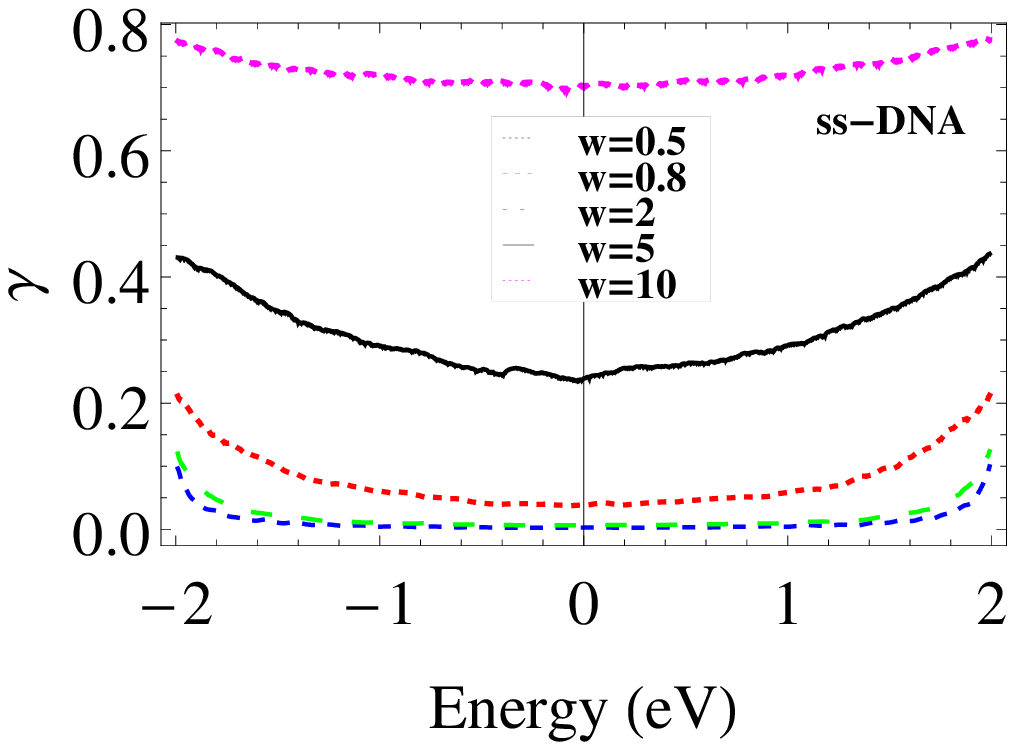}

  \end{tabular}

\caption{(Color online). Lyapunov exponent ($\gamma$) vs Energy 
for $n$=3 and $n$=5, model:2 and for ss-DNA at different disorder 
strength. Some specific features of the curve can be seen for $n$=5, 
and model:2. It is evident that ss-DNA being the most localized 
medium throughout the entire energy range.}

\label{fig3}
\end{figure}

 Fig.~\ref{fig3} presents variation of $\gamma$ with energy at different 
disorder strength for different models and molecules. For protein molecule 
with $n$=3, at low disorder system is less localized and $\gamma$ is close 
to zero. With increasing disorder, $\gamma$ increases and the system becomes 
more and more localized. However, at energies close to band centre (E=0 eV) 
localization is stronger than that at the band edges which is also evident from 
previous plots (Fig.~\ref{fig2}). For $n$=5, there is a typical shape in $\gamma$ within 
the energy region 1 to 1.5 eV (both -ve and +ve) which shows that at these energies 
system is maximally localized even at low disorder, with increasing disorder these 
features die out gradually. These specific features are results of their helical 
structure, the dispersion (E-k) of the molecule explicitly depends on the helical 
symmetry ($n$) (so the density of states also) and with changing $n$ specific features 
appear in different transport parameters. For model:2, there exists also the same 
feature at low disorder but near the band centre which vanishes with increasing disorder. 
This feature suggest that there is deficiency of accessible states around the band centre, 
with increasing disorder new states generates in these region at the cost of the other states 
around band edges and this feature vanishes. Another point to be noted that $\gamma$ is smaller 
in -ve energy region than +ve region, which indicates that the band is not symmetric with 
respect to energy and therefore effect of disorder is different on both side of E=0 eV. 
For ss-DNA, $\gamma$ increases with disorder, band edges are more localized than the band 
centre which is trivial in this case. 

\begin{figure}[ht]
\centering

  \begin{tabular}{c}

    \includegraphics[width=62mm]{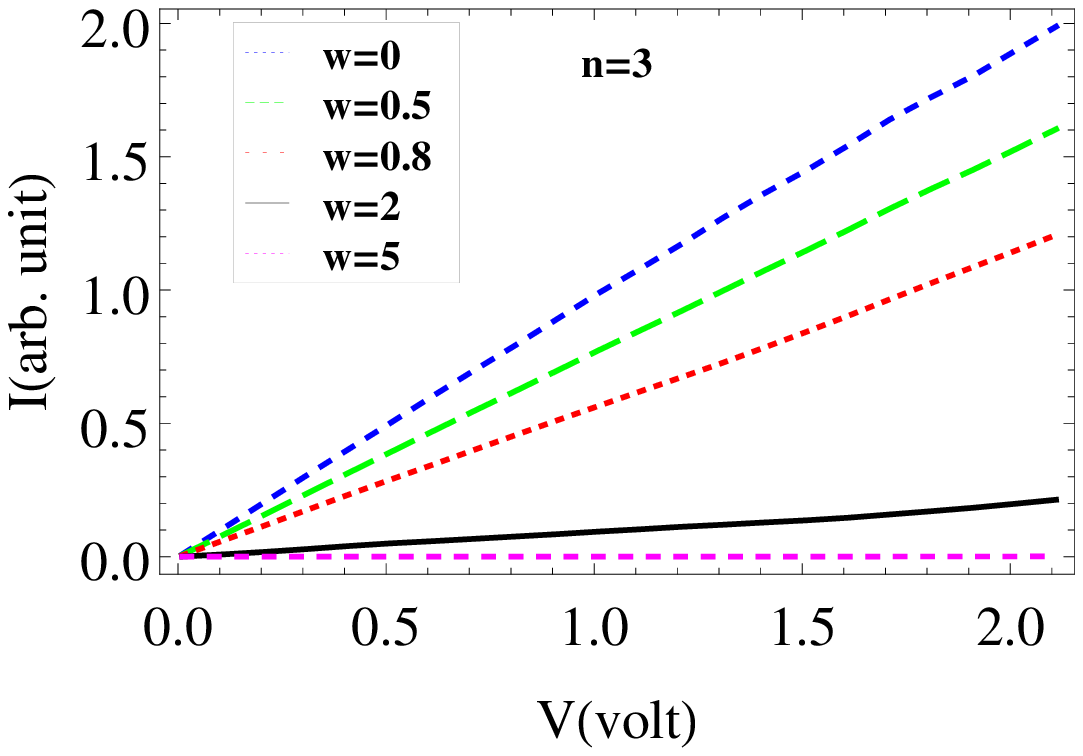}\\
   
    \includegraphics[width=60mm]{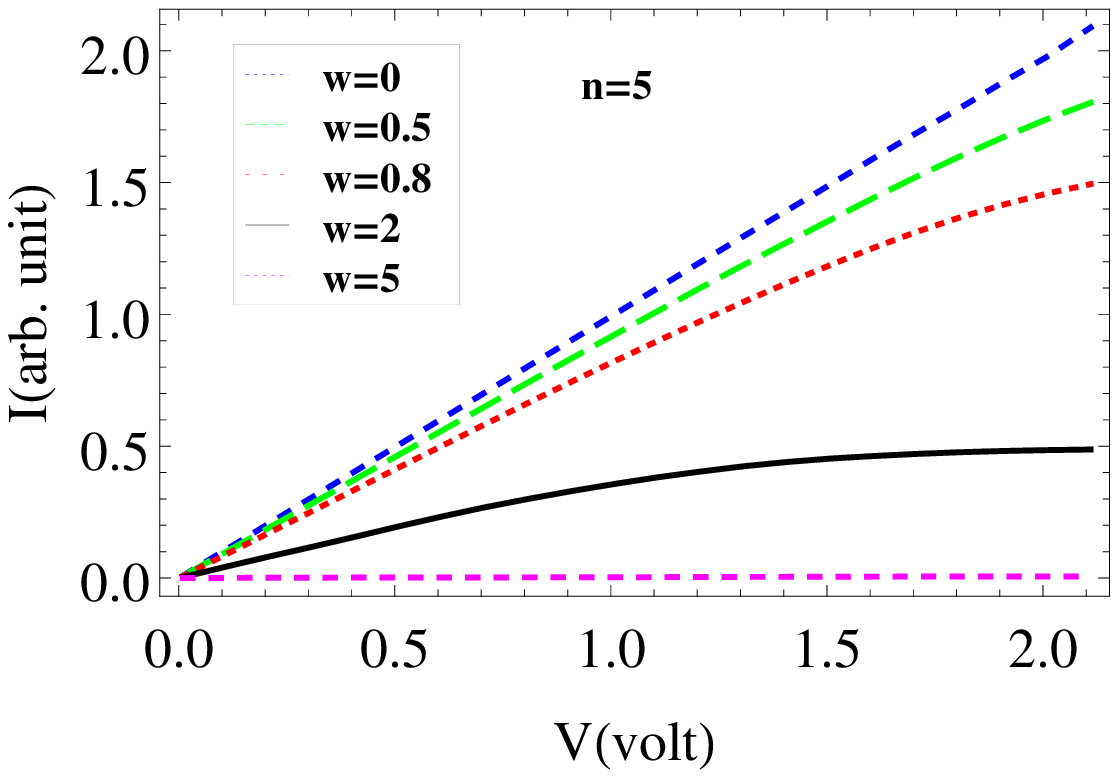}\\
    
    \includegraphics[width=60mm]{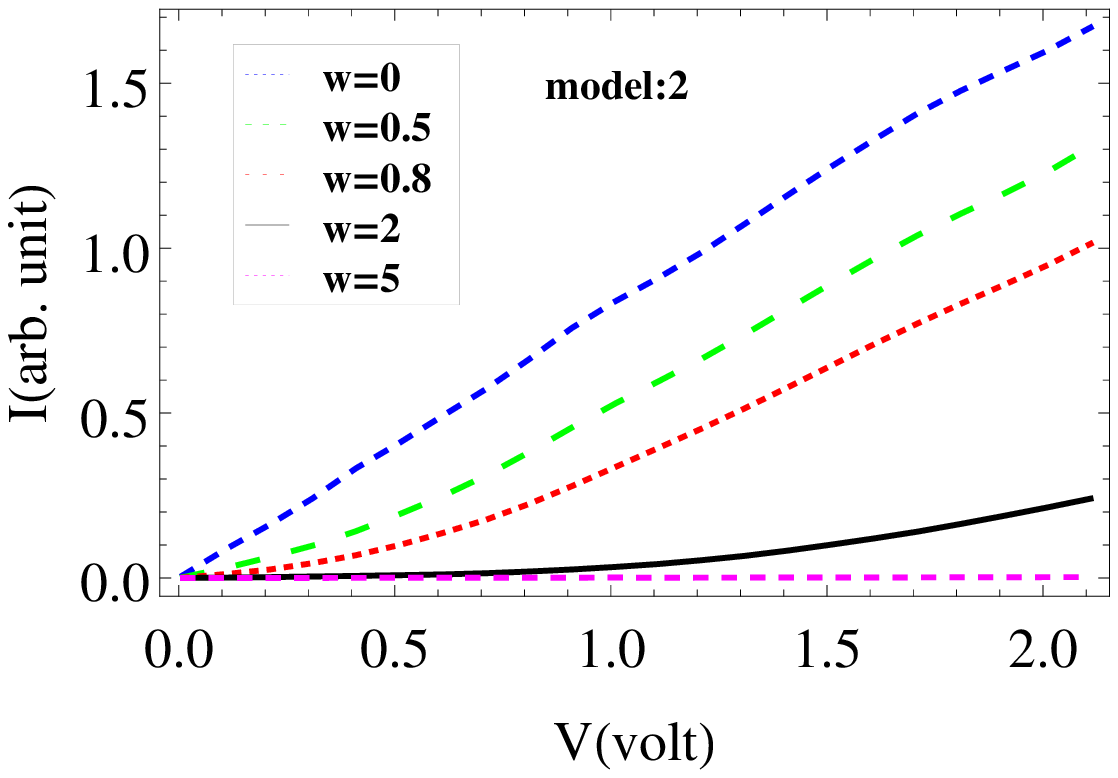}\\

    \includegraphics[width=62mm]{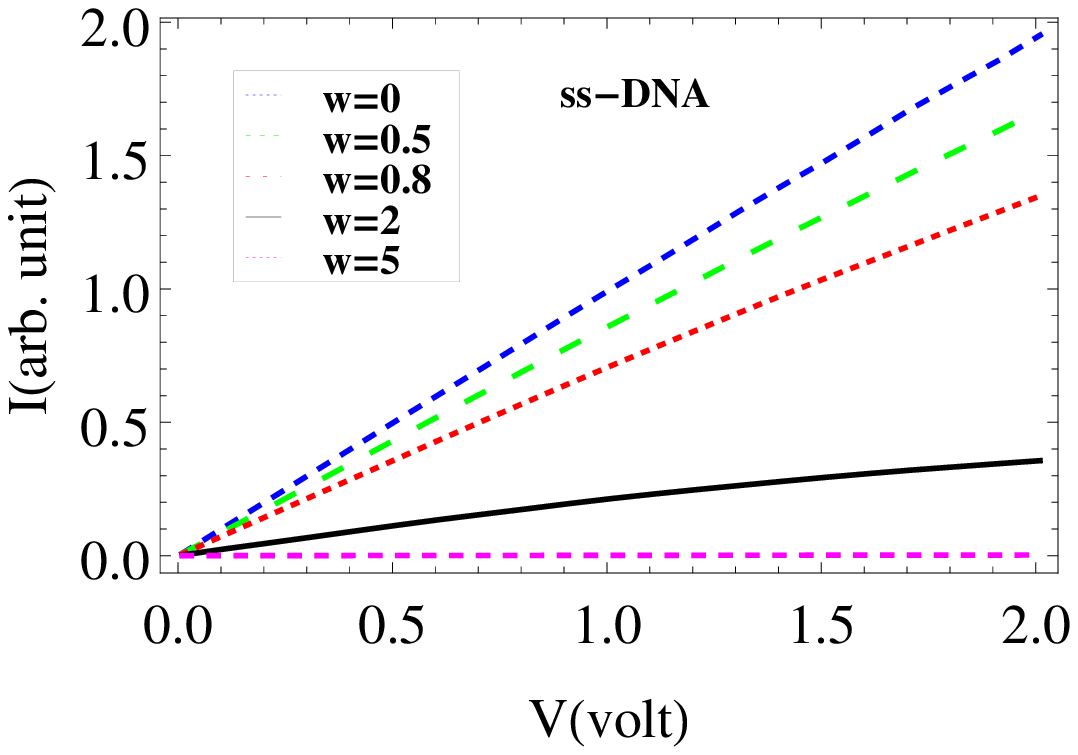}
    
  \end{tabular}

\caption{(Color online). I-V characteristics for different models and 
molecules. For w=0, all of the curves are nearly similar, with increasing 
disorder the responses become differentiable. It is clear from the 
plots that $n$=5 gives maximum current output under environmental 
fluctuations, that again shows it is the best possible conformation 
for the protein molecule for charge transport.}

\label{fig4}
\end{figure}

 In fig.~\ref{fig4} we show the I-V characteristics of the protein molecule using our 
proposed model for $n$=3 and 5. It is clear from the plots that for low disorder, current 
is almost linear and with increasing disorder (w=5) system becomes insulating. One 
difference is to be noted between $n$=3 and 5, current output is always greater 
for $n$=5 than $n$=3. This difference becomes more distinguishable under environmental 
fluctuations. If we increase $n$, electrons get more number of pathways to hop from 
one site to another. In other words, increasing $n$ provides a shortcut pathway for 
electrons to move along the helix from one pitch to the next bypassing all other amino 
acids in that pitch. Due to this galloping electrons encounter less scattering, hence 
the current increases. For $n$=5 at w=2, I-V response is linear at low bias but 
it saturates with increasing bias voltage which signifies that there are no more 
accessible states available to conduct electrons at higher bias. For model:2, at low disorder 
there is not so much variation in the current response with our model, but for appreciable 
disorder (w=2) response become semiconducting with cut-off voltage around 0.5 Volt, 
in contrast with linear response achieved in our model. This variation of current 
is due to the difference in these two models. The model:2 only caters the possibility 
of multiple charge conduction pathways but neglect the helical symmetry which is a 
fundamental structure of these biomolecules. Our model also incorporates this 
symmetry, due to which electron transport enhances and current output is always 
greater in our case ($n$=5). The I-V responses of the protein molecule under environmental 
effects match qualitatively well with experimental findings~\cite{jin,lee,ron}. 
For ss-DNA current output is quite similar with the protein molecule for clean case (w=0), 
but as soon as we apply disorder current decreases rapidly. If we compare ss-DNA with 
protein molecule ($n$=5), one can easily see that even under small disorder (w=0.5, 0.8) 
there is substantial drop in the current output for ss-DNA. This shows that single-helical 
proteins are much better conductor under external disturbances than ss-DNA. 

\section{conclusion:} 

 We present a detail analysis of electronic transport properties of single-helical 
protein molecules based on tight-binding framework. Though with recent progress in CISS single 
helical proteins are getting considerable attention these days but their basic electronic 
transport properties remain quite under-explored. We show that due to existence of 
both the helical symmetry and MCCP transport properties are less affected by external 
disturbances and these biomolecules can serve as a better alternative in nanoelectronics 
than its DNA counterparts (ss-DNA) or other artificial semiconductor nanowires 
which do not contain the possibility of MCCP. As a consequence of these features long 
range coherent electron transport can be possible in these biomolecules and effect of 
environment will also be lesser. 
We also show that by varying $n$ {\it i.e.}, by twist-stretching of the protein molecules 
one can get different responses (current outputs) under environmental effects. As an 
example, for $n$=5 system is more conducting than $n$=3. We compare our results 
with the model proposed by Guo et al.~\cite{guo_pnas}, that matches qualitatively well 
in almost every aspect. But as our model includes both MCCP and helical symmetry (both 
are actually present in single-helical proteins) it provides more number of conduction 
channels and system becomes less localized. Due to helical symmetry ($n$) the behaviour of 
localization properties also change which shows that using twist-stretching or conformal 
changes one can use a protein molecule in different ways. We check our results for other 
model parameters also {\it e.g.}, $l_c$=0.4 and 1.4 \AA~\cite{guo_pnas} but the 
qualitative behaviour remains the same. In summary our study shows 
the rigidity of transport properties of the single-helical protein molecules even under 
environmental effects due to presence of both the helical symmetry and MCCP, which 
promises that they can be used in future nanoelectronic devices with much more reliability.

\end{document}